\pdfoutput=1

\documentclass [11pt]{article}

\usepackage{amsmath,amsthm,enumitem,multicol,amssymb,mathrsfs,verbatim,graphicx,authblk}
\usepackage{bbm}
\usepackage{geometry,xcolor}
\usepackage[makeroom]{cancel}
\usepackage[normalem]{ulem}

\DeclareMathSymbol{\R}{\mathbin}{AMSb}{"52}

\renewcommand{\R}{\mathbb{R}}

\DeclareMathSymbol{\N}{\mathbin}{AMSb}{"4E}
\DeclareMathSymbol{\Z}{\mathbin}{AMSb}{"5A}

\begin{document}

\title{Aging a little: The optimality of limited senescence in \emph{Escherichia coli}}
 
\author{Natasha Blitvi\'{c}}
\affil{Department of Mathematics and Statistics\\ Lancaster University}

\author{Vicente I. Fernandez}

\affil{Department of Civil, Environmental and Geomatic Engineering\\ ETH Z\"urich}

\maketitle

\begin{abstract}

Recent studies have shown that even in the absence of extrinsic stress, the morphologically symmetrically dividing model bacteria \emph{Escherichia coli} do not generate offspring with equal reproductive fitness. Instead, daughter cells exhibit asymmetric division times that converge to two distinct growth states. This represents a limited senescence / rejuvenation process derived from asymmetric division that is stable for hundreds of generations. It remains unclear why the bacteria do not continue the senescence beyond this asymptote. Although there are inherent fitness benefits for heterogeneity in population growth rates, the two growth equilibria are surprisingly similar, differing by a few percent. In this work we derive an explicit model for the growth of a bacterial population with two growth equilibria, based on a generalized Fibonacci recurrence, in order to quantify the fitness benefit of a limited senescence process and examine costs associated with asymmetry that could generate the observed behavior. We find that with simple saturating effects of asymmetric partitioning of subcellular components, two distinct but similar growth states may be optimal while providing evolutionarily significant fitness advantages. 
\end{abstract}

\section{Introduction}

With the development of new experimental methods (including microfluidics \cite{rusconi2014microfluidics, zare2010microfluidic}, automated image analysis \cite{slack2008characterizing}, and imaging techniques \cite{li2012single, lechene2006high}) the recent decades have seen an explosion of microbial single cell data with large sample sizes. One consequence has been the realization that even clonal populations often demonstrate substantial variability in many characteristics or behaviors, termed ‘phenotypic heterogeneity’ \cite{ackermann2015functional, avery2006microbial}.  Phenotypic heterogeneity is an umbrella term that can encompass many types of heterogeneity, some carefully tuned by the organism through evolution (e.g. bet hedging) and others potentially unregulated and stochastic. 

A surprising result of this area of research was the observation of aging in morphologically symmetric \emph{Escherichia coli} \cite{stewart2005aging}. Senescence is the progressive deterioration of an individual’s fitness (through survival or reproduction rate) over time and is broadly occurring in multicellular organisms \cite{jones2014diversity, lemaitre2015early}.  As aging has been tied to physical differences between parent/offspring generations in which the parent retains older, damaged, material, it was assumed that unicellular organisms did not age \cite{moger2019microbial, kirkwood2000we}. Outside of some well-known examples (\emph{Saccharomyces cerevisiae} \cite{mortimer1959life}, \emph{Caulobacter crescentus} \cite{Ackermann1920}) with pronounced asymmetric division and observed senescence, unicellular microbes such as \emph{Escherichia coli} bacteria divide in an apparently morphologically symmetric manner. 

The aging of symmetrically dividing unicellular organisms is of particular interest because it may provide insight into the evolutionary origins of aging \cite{ackermann2007evolutionary}.  It is unlikely that the earliest organisms divided asymmetrically or had systems for rejuvenating offspring. Thus there are indications that the evolution of aging may be coupled to the development of asymmetric division \cite{ackermann2007evolutionary, chao2010model}.  In this context, the aging of \emph{Escherichia coli} which show parent/offspring differences in reproductive ability but are morphologically indistinguishable may elucidate this evolutionary transition.

In order to define an age structure among individuals in a population, the underlying mechanism in the aging of \emph{Escherichia coli} draws on the means by which the bacteria grow and divide.  \emph{Escherichia coli} are rod shaped bacteria with poles at either end.  During division, new poles are generated in the interior of the parent cell and each offspring contains an ‘old’ and ‘new’ pole.  In this way, the bacterium can be assigned an age through number of generations of the old pole.  Old poles in \emph{Escherichia coli} have been associated with asymmetric partitioning of protein aggregates with negative impacts on growth rates \cite{lindner2008asymmetric, winkler2010quantitative}. 

Since its initial discovery in 2005 \cite{stewart2005aging}, the study of senescence in \emph{Escherichia coli} has a revealed a more complex picture than originally expected. Subsequent works following parent cells for 200 generations did not observe progressive deterioration, instead reporting remarkably stable growth rates \cite{wang2010robust}. Subsequent work suggested that aging in \emph{Escherichia coli} was not an inherent part of the growth process, but a conditional response to external damaging stress \cite{rang2012ageing} including the fluorescent proteins used to visualize bacteria in some earlier experiments. Most recently, new experiments \cite{proenca2018age, lapinska2019bacterial} with unstressed \emph{Escherichia coli} have revealed a hybrid behavior.  Proenca et al. and {\L}api\'nska et al. both conclude that there is a persistent asymmetry in division times between offspring after the division of a parent. However this asymmetry does not lead to the progressive deterioration of senescence, but instead converges to two asymptotic growth states.  A parent cell in one of the two asymptotes divides into one offspring of the same growth rate and one that undergoes the aging/rejuvenation process towards the alternative asymptote over a period of four generations \cite{proenca2018age}.  These results are consistent with both the early observations of senescence (on shorter timescales) as well as the observations of long term stability in parent cells.

The impact of the two growth equilibria on the population fitness remains unclear. The limited aging and rejuvenation processes in the growth of \emph{Escherichia coli} results in the majority of individuals growing at either of these two equilibria.  This is a very different population structure than presumed in prior studies of bacterial senescence where the progressive accumulation of damage leads to the removal of an individual from the population \cite{chao2010model, ackermann2007evolutionary, chao2016asymmetrical}.  While Jensen’s inequality indicates that any variance in growth rates coupled with the convexity of exponential growth will lead to an increased population growth rate, the small differences in offspring doubling times (1-5\% \cite{wang2010robust, lapinska2019bacterial, proenca2018age, stewart2005aging}) may not be large enough to directly generate a meaningful fitness advantage.

In this work, we evaluate the impact of the recently characterized \emph{Escherichia coli} age structure on population growth using a novel combinatorial conceptual model.  The combinatorial model is naturally interpreted as a generalized Fibonacci recurrence, providing intuitive links between the total population size and that of the bacteria growing at the two equilibria. We subsequently use the model to examine the cost in generating offspring asymmetry and potential differences between stressed and unstressed states. Analytical models for heterogeneous growth have been developed and utilized in many microbial contexts, from understanding the noisy growth of bacterial colonies \cite{bellman1948theory, kendall1952choice} to predicting growth rates of the inherently heterogeneous budding yeast \cite{Olofsson2011, spears1998asymmetric}.  Such models focus on situations with large discrepancies in growth rates and cannot be applied to bacteria such as \emph{Escherichia coli}. We demonstrate that the stable small asymmetry in growth rates in unstressed \emph{Escherichia coli} can produce significant fitness advantage over completely symmetric division. 

\section{A Combinatorial Model of Bacterial aging with Two Equilibria}
\label{sec-combinatorial}

\begin{figure}\includegraphics{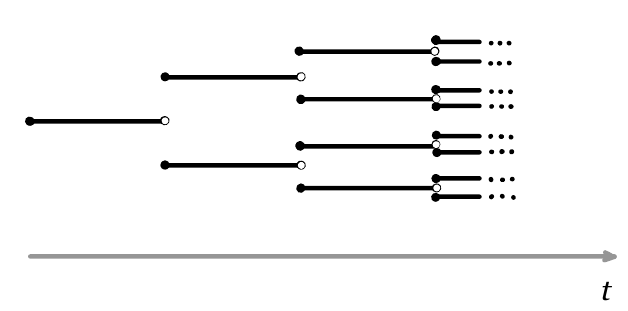}\includegraphics{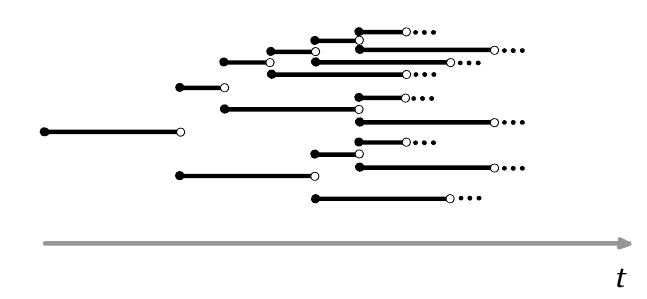}\centering\caption{(a) Bacterial division under equal lifetimes. (b) Bacterial division with a $3:1$ asymmetry in daughter cell doubling times.}\label{fig-tree}\end{figure}

Consider the standard model where a bacterium is born at time 0 and at multiples of a fixed doubling time $T_0$, each bacterium divides into two. At any time $t$, there are therefore $2^{\lfloor t/T_0\rfloor}$ bacteria present and the evolution of the process can be can be represented as a standard binary tree (Figure~\ref{fig-tree}A). We expand this model to treat aging that is consistent with the two growth equilibria of Proenca et al. \cite{proenca2018age}. Namely, we consider asymmetric division with consistent but small discrepancies in doubling time in the daughter cells (Figure~\ref{fig-tree}B), where each division results in one ``faster growing" and one ``slower growing" cell. This approach simplifies the four generation transition between equilibria \cite{proenca2018age} while capturing the central age structure of the population. 

\subsection{Population growth rate}

For the combinatorial model of bacterial aging, we will fix a time increment $\delta>0$ and consider the bacterial doubling times $\delta m$ and $\delta k$, where $k$ and $m$ are some fixed integers with $1\leq k< m$. Here, $k$ and $m$ are assumed to be relatively prime, as any common factors can be absorbed into the time increment $\delta$. Since every bacterium has one of two doubling times, the discretization of time allows us to apply tools from combinatorics and linear difference equations without introducing any approximations.

Initially, a bacterium is born at time 0 and undergoes its first division at time $\delta m$. Then, starting with this division and for all divisions thereafter, each bacterium in the system splits into a child that will undergo division at time $\delta k$ and a child that will undergo division at time $\delta m$. We  denote by $a_n^{(k,m)}$, for $n=0,1,2,\ldots$, the total number of bacteria at time $t\in [n\delta,(n+1)\delta)$. We also track the number of simultaneous divisions occurring at any time $\delta n$ by $d_n^{(k,m)}$ ($n=0,1,2,\ldots$).

After the initial division (for $n\geq m+1$) the divisions can be described by a recurrence relation
\begin{equation}d^{(k,m)}_n=d^{(k,m)}_{n-k}+d^{(k,m)}_{n-m},\label{eq-recurrence-d}\end{equation}
with $d^{(k,m)}_0=\ldots=d^{(k,m)}_{m-1}=0$ and $d^{(k,m)}_{m}=1$ \label{lemma-d}.  (See Supplemental Materials.)   The total number of bacteria in the population follows a similar recurrence relation for $n\geq m$, 
\begin{equation}a_n^{(k,m)}=a_{n-k}^{(k,m)}+a_{n-m}^{(k,m)},\label{eq-recurrence}\end{equation} but with differing initial conditions where $a_0^{(k,m)}=\ldots=a_{m-1}^{(k,m)}=1$.

These recurrence relations have the form of a \emph{generalized Fibonacci recurrence}
\begin{equation}g_n=g_{n-k}+g_{n-m},\tag{GF}\label{eq-GF}\end{equation}
where $k$ and $m$ are relatively prime natural numbers. Letting  $k=1$ and $m=2$ recovers the well-known \emph{Fibonacci numbers} (initial conditions $g_0=0, g_1=1$) and \emph{Lucas numbers}  (initial conditions $g_0=2, g_1=1$). Both of these sequences have been extensively studied and satisfy a wealth of mathematical properties \cite{fibonacci-knott,Vajda,Koshy}, often unexpectedly featuring in various areas of mathematics and natural sciences. Models of asymmetric yeast division have also been based on a similar recurrence with $k=1$ \cite{Olofsson2011}, but the generalized recurrence with integer $1\leq k\leq m$ is required to capture the much smaller growth asymmetries generated by bacteria with morphologically symmetric division.

By expanding the recurrence relation \eqref{eq-recurrence} in terms of the sequence of divisions $d_n^{(k,m)}$, we find (see Supplemental Materials) that the closed-form solution to the number of bacteria at time $n\delta$ is  
\begin{equation}a_n^{(k,m)}=\sum_{r}\frac{(1-r^k)}{(1-r)(mr^{m}+kr^{k})}\,\frac{1}{r^{n}},\label{eq-explicit}\end{equation}
where the sum is taken over all (complex) roots $r$ of the characteristic equation
\begin{equation}1-r^k-r^m=0.\tag{CE}\label{eq-poly}\end{equation}

For all $1\leq k<m$ ($k,m\in\mathbb N$), \eqref{eq-poly} has a unique positive root $r_0$, which is such that $|r|> r_0$ for all other roots $r$ of \eqref{eq-poly}.  As a result, $r_0$ dominates the asymptotic growth of the sequence. That is, for large $n$, 
\begin{equation}a_n^{(k,m)}\sim \frac{(1-r_0^k)}{(1-r_0)(mr_0^{m}+kr_0^{k})}\,r_0^{-n}.\label{eq-asymptotic}\end{equation}

Comparing \eqref{eq-asymptotic} to an exponential population growth of the form $2^{{\mu}t}$ with growth rate $\mu$, we see that the growth rate with heterogeneous division is $-\delta^{-1}\log_2(r_0)$. The root $r_0$ is also the ratio of consecutive terms in \eqref{eq-explicit}, $a_n^{(k,m)}/a_{n+1}^{(k,m)}$, ensuring a positive growth rate.  We can further bound the population growth rate (Supplemental Materials), finding that 
\begin{equation}
\frac{1}{m}<\frac{2}{k+m}<-\log_2(r_0)<\frac{1}{k}.\label{eq-inequality}
\end{equation}
As expected, for large $n$, the population with two different growth states grows faster than a population with only the larger doubling time and slower than an analogous population with the smaller doubling time.  However, \eqref{eq-inequality} also indicates that the heterogeneous growth is strictly faster than the growth of a population whose doubling time is average of the two doubling times, $(k+m)/2$, confirming the intuition derived from Jensen's inequality. More importantly, by numerically evaluating $r_0$ as the only real positive root of equation \eqref{eq-poly}, we can quickly compute the exact large-time population growth rate and quantify the effect of heterogeneity on the speed of population growth.

\subsection{Composition of the population}
\label{sec-composition}
We now turn to the problem of characterizing the composition of the bacterial population at time step $n$ as differentiated by the lifetime. Namely, let $b^{(k,m)}_n$ $(n=0,1,2,\ldots)$ denote the number of bacteria alive at time $t\in [n\delta,(n+1)\delta)$ having \emph{doubling time $\delta k$}. Similarly, let $c^{(k,m)}_n$ denote the number of bacteria alive at time $t\in [n\delta,(n+1)\delta)$ having \emph{doubling time $\delta m$}. 
Both sequences satisfy the same recurrence as $(a^{(k,m)}_n)_n$ and $(d^{(k,m)}_n)_n$, starting from a different initial condition. Namely, for $n\geq k+m$, we have
\begin{eqnarray}b^{(k,m)}_n&=&b^{(k,m)}_{n-k}+b^{(k,m)}_{n-m}\label{eq-recurrence-b}\\
c^{(k,m)}_n&=&c^{(k,m)}_{n-k}+c^{(k,m)}_{n-m}\label{eq-recurrence-c}
\end{eqnarray}
with 
$$b^{(k,m)}_0=\ldots=b^{(k,m)}_{m-1}=0\quad \text{and} \quad b^{(k,m)}_{m}=\ldots=b^{(k,m)}_{m+k-1}=1$$
$$c^{(k,m)}_0=\ldots=c^{(k,m)}_{m+k-1}=1.$$
(Recall that $1\leq k<m$.) A brief proof is included in the supplementary materials.  

Interestingly, the two sequences counting each bacterial subgroup are shifted versions of the sequence for the total population number.  Indeed, $b_n^{(k,m)}=a_{n-m}^{(k,m)}$ for $n\geq m$, and $c_n^{(k,m)}=a_{n-k}^{(k,m)}$ for $n\geq k$. This allows one to straightforwardly derive closed-form expressions for large time using the results for the total population. Namely, for $n\geq k$, 
\begin{eqnarray}b^{(k,m)}_n&=&\sum_{r}\frac{(1-r^k)^2}{(mr^{m}+kr^{k})(1-r)}\,\frac{1}{r^{n}},\label{eq-explicit-b}\\
c^{(k,m)}_n&=&\sum_{r}\frac{r^{m+k}}{(mr^{m}+kr^{k})(1-r)}\,\frac{1}{r^{n}},\label{eq-explicit-c}
\end{eqnarray}
where the sum is taken over all roots $r$ of \eqref{eq-poly} $1-r^k-r^m=0$. When $k$ and $m$ are relatively prime, for large $n$, 
\begin{eqnarray}b_n^{(k,m)}&\sim& \frac{(1-r_0^k)^2}{(mr_0^{m}+kr_0^{k})(1-r_0)}\,r_0^{-n},\label{eq-asymptotic-b}\\
 c_n^{(k,m)}&\sim& \frac{r_0^{k+m}}{(mr_0^{m}+kr_0^{k})(1-r_0)}\,r_0^{-n},\label{eq-asymptotic-c}
\end{eqnarray}
where $r_0$ is again the unique positive root of \eqref{eq-poly}, as in \eqref{eq-explicit}.

A natural question concerns the relative proportions of bacteria, among all those alive at some specified (large) time $t$, with a given doubling time $\delta k$ (resp.\ $\delta m$).  Note that the total population as well both subgroups grow exponentially with the same growth rate ($-\log_2(r_0)$). Na\"ively, as each division produces one child of lifetime $\delta k$ and one child of lifetime $\delta m$, one might expect the proportions of the slower dividers to the faster dividers to be even and the aforementioned ratio to be asymptotically equal to $1/2$. This argument fails to take into account the fact that bacteria with longer lifetimes contribute to the population for a longer time before dividing and are therefore more heavily represented among the population existing at any fixed time. If one considers the situation where one of the two daughter cells is a spore with infinite doubling time, the population would quickly be dominated by such spores despite the growth being entirely due to the non-spore daughter cells. The exact proportions of both subgroups take a surprisingly simple form:

\begin{equation}\lim_{n\to\infty}\frac{b_n^{(k,m)}}{a_n^{(k,m)}}=r_0^m,\quad\quad \lim_{n\to\infty}\frac{c_n^{(k,m)}}{a_n^{(k,m)}}=r_0^k.\end{equation}
Since $k<m$, the proportion of bacteria with doubling time $k$ is between $0$ and $1/2$, where we approach the lower bound by letting $k$ be small and $m$ large, as in the previous example.  The upper bound is reached with $m$ large as $k/m$ approaches $1$.

\subsection{Robustness of the combinatorial model}
\label{sec-robustness}

The model based on the generalized Fibonacci recurrence \eqref{eq-GF} is `noiseless', relying on each division to produce one child with doubling time of exactly $\delta k$ and another with a doubling time of exactly $\delta m$. This simple formulation is suited for exploring fundamental questions in bacteria with two equilibrium growth states, but may raise questions on the validity of the model when the assumptions are not strictly adhered to. We find that the combinatorial aging model applies more generally when daughter cells have small stochastic variability in the division times. Extending the methods of Bellman and Harris \cite{bellman1948theory}, we can describe the behavior of a generalized branching process, where each division engenders one child with \emph{mean lifetime} $\delta k$ and another with \emph{mean lifetime} $\delta m$. (See Supplemental Materials.) Under sufficient regularity conditions, the population almost always grows with rate $\alpha$, determined by the following fixed-point equation:
\begin{equation}
1-\int_0^\infty e^{-\alpha t} dF_1(t)-\int_0^\infty e^{-\alpha t} dF_2(t)=0,
\tag{FP}\label{eq-fp}
\end{equation}
where $F_1$ is the cumulative distribution function of the lifetime of the first child and $F_2$ is that of the second child. If the two children have fixed lifetimes of $\delta k$ and $\delta m$ respectively, \eqref{eq-fp} reduces to \eqref{eq-poly}, where $r=e^{-\alpha}$.  Furthermore, it can be shown (see Supplementary Materials) that as $F_1$ and $F_2$ become progressively more concentrated around their means, such as when the lifetimes are uniform on $[k-\epsilon,k+\epsilon]$ and $[m-\epsilon,m+\epsilon]$ (with vanishing $\epsilon>0$), the root of \eqref{eq-fp} converges to the unique positive root of \eqref{eq-poly}. In other words, the combinatorial model is not `rigid', but rather provides a realistic approximation whose accuracy increases as the noise variance decreases.

\subsection{Asymptotics for small division asymmetries}
Finally we take a closer look at some analytic properties of the unique positive root of the characteristic equation $1-r^k-r^m=0$, governing the asymptotics of equations \eqref{eq-asymptotic}, \eqref{eq-asymptotic-b}, and \eqref{eq-asymptotic-c}. We are interested in behavior of this root as the difference between $k$ and $m$ grows.  Let us consider a bacterial population where daughter cells have discretized doubling times $k=\ell_0-x$ and $m=\ell_0+x$, where $\ell_0$ is the discretized homogenous doubling time. In this scenario, $x$ is a measure of the strength of the asymmetry between offspring. Under this formulation, the population growth rate can be expressed as (see Supplemental Materials): 
\begin{equation}
 {\mu}(x)\delta=\frac{1}{\ell_0}+\frac{\log(2)}{2\ell_0}\left(\frac{x}{\ell_0}\right)^2+O\left(\frac{x}{\ell_0}\right)^4.\label{eq-quadratic}
\end{equation}
For small $x/\ell_0$, the growth rate increases quadratically with doubling time heterogeneity.

\section{Results}
\subsection{Quantification of growth benefits}
\label{sec-biological-implications}

Despite the integer formulation, the combinatorial aging model can be utilized to quickly explore the impact of bimodal growth heterogeneity with nearly arbitrary doubling times, including the recently characterized behavior of unstressed \emph{Escherichia coli} with two growth equilibria \cite{proenca2018age}.  After the growth equilibria are converted into the framework of the model, one only needs to numerically calculate the unique positive root of the characteristic equation \eqref{eq-poly} (between known bounds from Eq.\eqref{eq-inequality}) in order to describe the long-term growth rate and community composition of the bacterial colony. 

A symmetrically dividing population with doubling time $T_0$ (growth rate $1/T_0$) will be used as a reference for bacterial populations with asymmetric division. In this scenario, the individual doubling times and growth rates are the same as that of the population as a whole. For asymmetrically dividing bacteria, we quantify the magnitude of the asymmetry by $\gamma \in (0,1)$ which represents a deviation from the symmetrically dividing case $\gamma = 0$. 

We first consider heterogeneity under the assumption that an asymmetric division will increase the doubling time of one daughter cell by the same amount that it decreases the doubling time of the other (a zero-sum tradeoff). 
The division times of the two offspring are specified by $k=(1-\gamma)T_0/\delta$ and $m=(1+\gamma)T_0/\delta$ where $\delta$ is again a time interval. In this formulation, the average doubling time of the two offspring remains equal to the symmetric division case, regardless of the asymmetry ($\gamma$). In practice, $\gamma$ and $T_0$ will be truncated so $\delta$ can be chosen such that $k$ and $m$ are both integers and relatively prime, meeting the conditions for the combinatorial model.

\begin{figure}\centering
\includegraphics[scale=0.6]{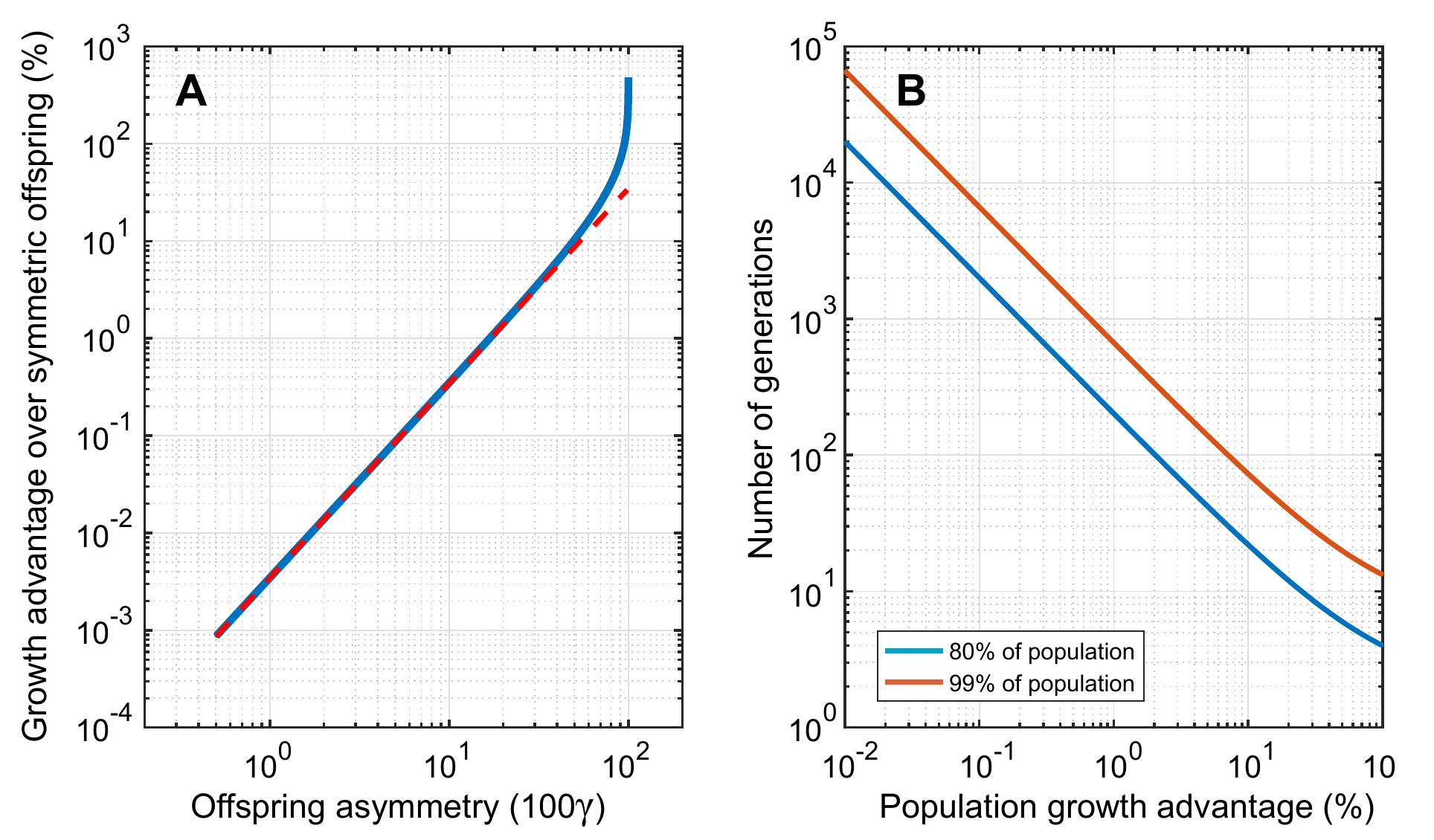}
\caption{Population growth advantage under dual-equilibrium age structure for bacteria.  (A) As the division asymmetry increases under a linear tradeoff between offspring, the overall growth rate of a bacterial colony (blue) increases with respect to the homogenous case. The increase in growth is unbounded as the offspring asymmetry approaches 100\% as the doubling time of one daughter cell approaches zero. The red dashed line depicts the analytical relationship in the vicinity of zero asymmetry (Eq. \eqref{eq-quadratic}).  (B) Small differences in population growth rates can quickly lead to dominance between competing strains. The blue and red curves indicate the number of generations for a faster growing population to reach $80\%$ (and $99\%$ respectively) of the combined population when both start at the same size.  The $x$-axis describes the prescribed difference in growth rates between the two competing populations.}
\label{fig-X}
\end{figure}

Under this assumption of a direct tradeoff between the progeny's doubling times, the population growth rate monotonically increases with increasing asymmetry in division (Figure~\ref{fig-X}A). In general, the increase in the population growth rate is modest for small differences in doubling time among daughter cells. For the $1$ to $5\%$ asymmetry observed across different single-cell experiments \cite{wang2010robust, lapinska2019bacterial, proenca2018age, stewart2005aging}, the increase in population growth rate ranges from $0.003$ to $0.09\%$ above the symmetrically dividing case. At an asymmetry of $17\%$, the increase in growth rate first surpasses $1\%$. For $\gamma$ less than $14\%$, the increase in population growth rate compared to the symmetric division reference case is proportional to $\gamma^2$ with an error $<0.01\%$, indicating that the analytical expansion around $\gamma = 0$, that is, \eqref{eq-quadratic} with $\gamma = x/\ell_0$, applies for a fairly broad range. 

The small differences in population growth rate associated with $1-5\%$ doubling time asymmetry in daughter cells are unlikely to cause short term benefits in bacterial populations, such as a competitive advantage in colonizing a new resource. However, they may have  significant impacts at the ecological level due to the exponential growth. To illustrate the impact of small growth rate differences, consider two bacterial populations that populate the same environment and initially have the same population size. If one population grows $1\%$ faster than the other, that population will represent $80\%$ of the total amount of bacteria in $200$ generations (Figure~\ref{fig-X}B). With a $0.1\%$ growth rate advantage, it would take $2000$ generations. For a generation time between $0.5$ and $1$ hour, this would correspond to a period of approximately one to three months of competition. A consistent growth advantage of this magnitude can be significant on evolutionary timescales, particularly for bacteria which have relatively short generation times. 

\subsection{Costs of asymmetry}
In bacteria, it is unlikely that asymmetric division leads to zero-sum tradeoffs in division times unless the asymmetry is very small. Linear relationships between the quantity of specific subcellular components (such as damage products) partitioned into offspring and the fitness of those cells have been commonly used in bacterial aging models \cite{chao2016asymmetrical,chao2010model,ackermann2007evolutionary}, in the absence of information on the specific underlying mechanisms.  

\begin{figure}\centering
\includegraphics[scale=.65]{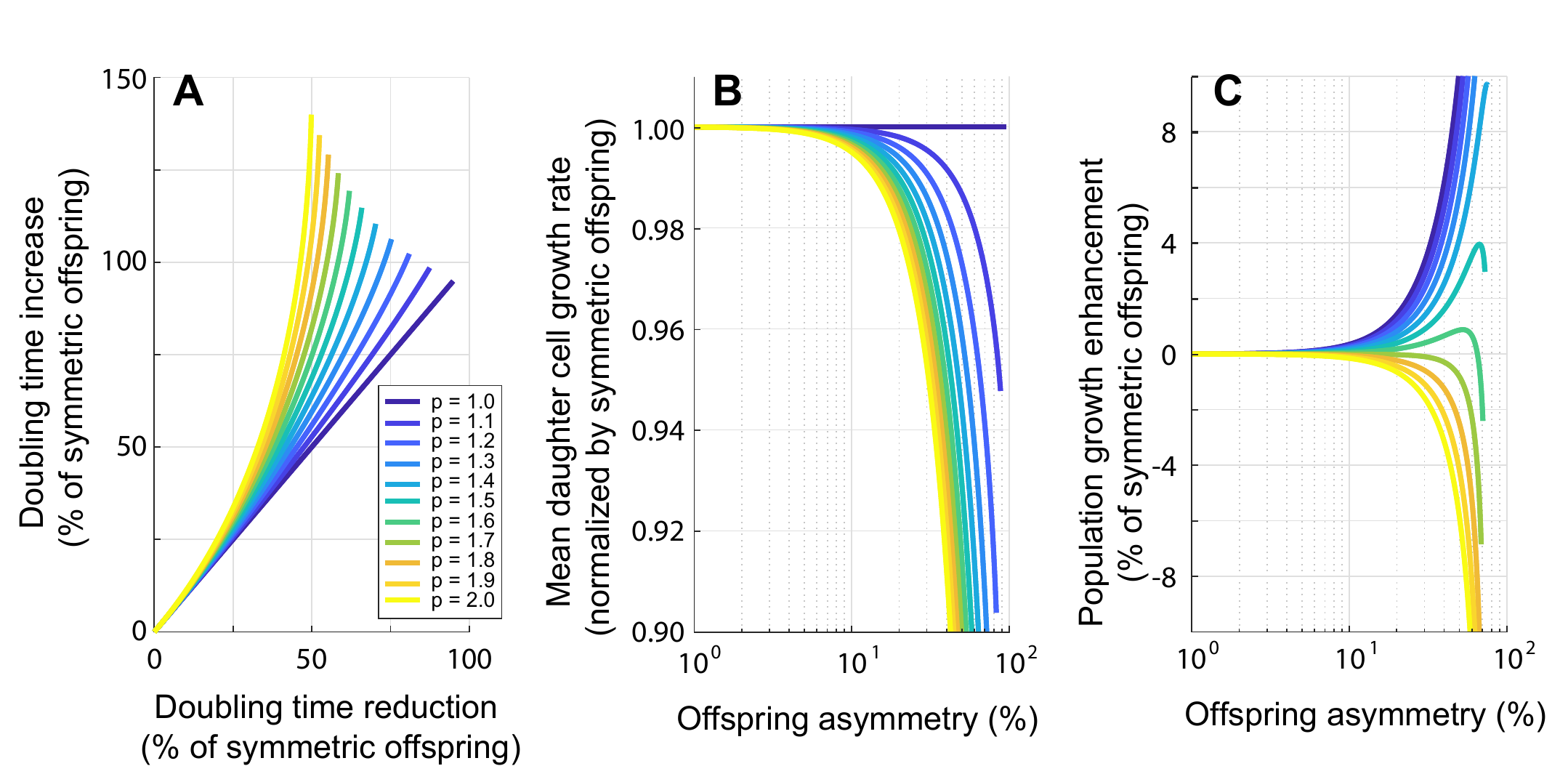}
\caption{Population growth under different tradeoffs in generating offspring asymmetry.  (A) Under different tradeoff models, the growth rate benefit of the rejuvenated offspring saturates ($x$ axis), while the cost to the aged cell continues to increase ($y$ axis). Different tradeoff curves are plotted following equation \eqref{eq-tradeoff}, including the case in which there is no cost to asymmetry $(p=1)$. Each curve depicts valid combinations of $k$ and $m$ for $0<\gamma<1$. (B)  When $p>1$, any asymmetry is coupled with an effective reduction in the combined offspring growth rate. The offspring asymmetry here refers to the resultant difference in doubling times, not the underlying asymmetrical partition ($\gamma$). All curves are for $0<\gamma<1$. (C) Increasing costs in generating offspring asymmetry sharply reduce the growth benefit of heterogeneity to the population relative to a population with symmetric division.  For intermediate levels ($p = 1.4- 1.7$) there is an optimum asymmetry smaller than 100\% for maximizing the population growth enhancement. The offspring asymmetry here refers to the difference in doubling times resulting from $0<\gamma<1$, not the underlying asymmetrical partition ($\gamma$).}\label{fig-Y}
\end{figure}

Proteins and other components in bacteria that are unevenly divided are likely to have an indirect impact on doubling times with some predictable nonlinearities. For a given set of environmental conditions, it is safe to assume that there is a lower limit in doubling time associated with the uptake of resources and production of biomass needed to divide. As a result, the benefit (shorter doubling times) to one cell in asymmetric division is likely to saturate or have diminishing returns as a function of the asymmetry. In contrast, it would be possible for a cell to extend their doubling time indefinitely and essentially cease to grow due to the accumulation of products with a negative impact on fitness. We will assume that the cost (increased doubling time) to an offspring under asymmetric division accelerates with increasing asymmetry. This corresponds to the idea that a cell may repair or compensate for a small amount of damage, but as the damage increases and potentially displaces undamaged components, the cell function is increasingly impacted. To represent these relationships explicitly, we let 
\begin{equation}
 \frac{k\delta}{T_0}=\frac{1}{p}\left((1-\gamma)^p-1\right)+1,\quad \frac{m\delta}{T}=\frac{1}{p}\left((1+\gamma)^p-1\right)+1 \label{eq-tradeoff}
\end{equation}
where $\gamma$ is the asymmetry at division as defined earlier, $T_0$ is the symmetric division doubling time (recovered when $\gamma=0$), and $p$ is a parameter for the strength of the nonlinearity. The tradeoff is plotted in Figure~\ref{fig-Y}A for several values between $1\leq p\leq 2$. When $p = 1$, the previous model with direct tradeoffs between daughter cells is recovered. When $p > 1$, the model produces saturating benefits and accelerating costs. At $p = 2$, this results in a maximum doubling time reduction of $50\%$ and maximum increase of $140\%$. The slightly complex form of \eqref{eq-tradeoff} is necessary to maintain the asymptotic behavior of the direct linear tradeoff near $\gamma=0$. In this way, for very small asymmetry $\gamma$, the behavior matches the original zero-sum tradeoff regardless of $p$, and a fair comparison between the cases can be made.  

An alternative perspective on the costs of aging and asymmetric division is that generating the asymmetry, whether by differentially partitioning damage from the parent cell or some other mechanism, requires resources that detract from the growth of the parent and subsequently both offspring. The saturating and accelerating relationships between asymmetry and doubling time discussed in the previous paragraph effectively create a similar outcome. Since the reward to one offspring cell is smaller and the cost to the other is larger, increasing the difference between offspring doubling times also decreases the average growth rate of both daughter cells combined when $p > 1$ (Figure~\ref{fig-Y}B). This can be interpreted as a cost to creating asymmetry in offspring.  

With the introduction of a nonlinear relation between offspring fitness and asymmetric partitioning as previously outlined, or equivalently a nonlinear cost to generating asymmetry, the growth advantage from large asymmetric division is substantially reduced.  Increasing the cost of asymmetric division, $p$, strictly reduces the growth advantage at all asymmetries $\gamma$, but the impact is largest for large asymmetry. This has the effect of creating a local maximum at intermediate levels of division asymmetry when $p$ is between  $1.4$ and $1.7$. As $p$ increases in this range, the position of the local maximum crosses from the outer extreme ($\gamma = 1$) to symmetric division ($\gamma = 0$).  For larger $p$, the local maximum is so close to $\gamma = 0$ that effectively any asymmetry in division reduces the population growth rate. 

\subsection{Community composition and observation}

\begin{figure}\centering
\includegraphics[scale=0.6]{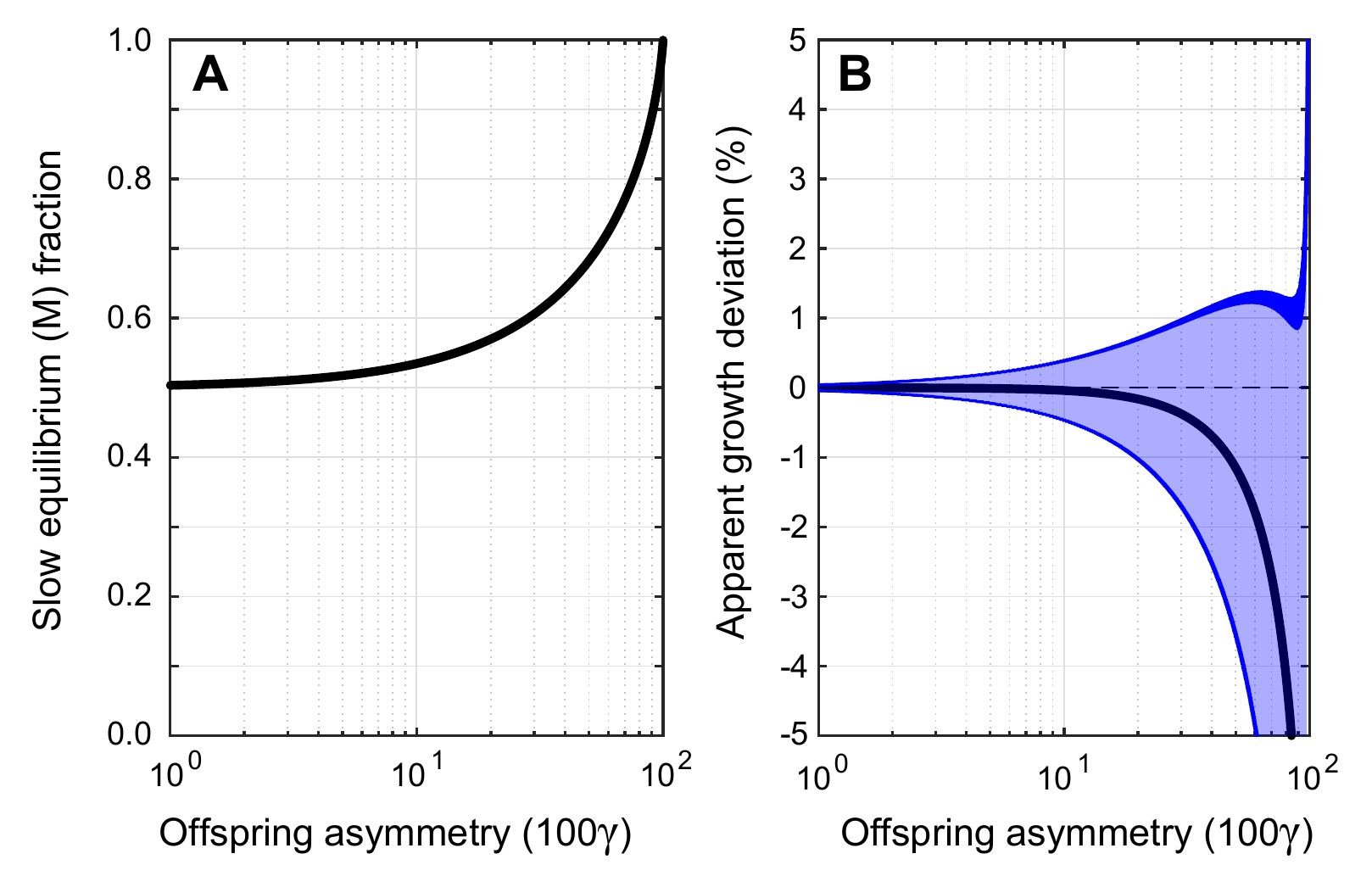}
\caption{Population composition in a zero-sum heterogeneous community. (A). As the heterogeneity between daughter cells increases, slow growing cells become an increasing fraction of the community. (B). The average growth rate of an instantaneous population (black line) always underestimates the true population growth rate (dashed black line), increasing with heterogeneity.  If a single cell experiment uses one thousand cells to calculate the growth rate, the estimated growth rate will have an 80\% chance to fall between with blue curves.}
\label{fig-Z}
\end{figure}

As the discrepancy in doubling times between progeny ($\gamma$) grows, the community composition at any particular time becomes skewed towards slow-growing individuals (Figure~\ref{fig-Z}A). This occurs the despite the fact that fast and slow-growing cells are always created in conjunction and in equal numbers, because the slower growing cells persist longer before dividing. As mentioned in \ref{sec-composition}, intuition is provided by the example of sporulating bacteria where the population quickly dominated in abundance by inactive cells. For the degree of growth asymmetry generated by \emph{Escherichia coli} without external stress, the population composition remains approximately evenly divided between the two growth equilibria (slow growing cells form $<51.7\%$ of total population).

As single cell experimental techniques have enabled direct measurements of the distribution of characteristics in populations of bacteria \cite{wang2010robust, kiviet2014stochasticity, taheri2015cell, hashimoto2016noise}, experimentalists should be wary of composites of individual measurements that may provide erroneous assessments of the population. In practice, a common estimate of the population growth rate is the mean growth rate averaged among all observed individuals (e.g. \cite{taheri2015cell} and many others).  This apparent mean estimator provides a better estimate of the population growth rate than a homogenous model based on the average rate of the two daughter cells $(1/T_0)$. However, the estimator based on individual growth rates will consistently underestimate the population growth rate (Figure~\ref{fig-Z}C). For small degrees of heterogeneity (less than $\sim 45\%$ asymmetry), this underestimation bias is less than $1\%$ of the true population growth rate. However, as the discrepancy between daughter cells increases further, as it might under higher extrinsic stress, the magnitude of the underestimate grows quickly. In addition, given the significant contribution to the population growth rate from an increasingly small fraction of the bacterial population, measurements of the population growth rate from a finite number of individual bacteria will have significant error. In Figure~\ref{fig-Z}C, the apparent mean estimator based on one thousand random chosen cells will fall into the shaded blue region with an $80\%$ likelihood. 

If the primary source of bacterial heterogeneity in an experiment is due to division asymmetry between two consistent growth rates, or any other situation which leads to a population with two growth equilibria, a more accurate experimental approach for estimating the population growth rate would be to identify the two doubling times directly from the bimodal distribution of individual cell doubling times.  These could then be directly used to identify $k$, $m$ and $\delta$ and estimate the population growth rate by solving the characteristic polynomial \eqref{eq-poly}.

\section{Discussion}
In the recent works by Proenca et al.\cite{proenca2018age} and {\L}api\'nska et al.\cite{lapinska2019bacterial}, it was shown that \emph{Escherichia coli} under unstressed conditions converge to persistent asymmetry during division that creates two dominant growth rates in the population. These observations resolve some of the conflicting observations that followed the initial observation of aging in \emph{Escherichia coli} by showing that initially progressive senescence will plateau at fast and slow growing equilibria. While there is now a cohesive explanation for the observed aging behavior, it remains unclear why these two growth state equilibria exist and \emph{Escherichia coli} exhibit this `limited' senescence. In principle, bacteria with old poles could continue to accumulate all growth-limiting subcellular components during division even in relatively unstressed conditions, benefiting from the previously studied resulting fitness advantage \cite{chao2010model,ackermann2007evolutionary}.

Via the explicit combinatorial model presented in this manuscript, we have demonstrated that a small but stable asymmetric partitioning could be optimal over long multi-generational timescales. Factoring in saturating benefits and compounding costs that are likely to impact the relation between asymmetric partitioning of damage and offspring creates an optimal intermediate asymmetry that can be moved arbitrarily close to the symmetric division state by increasing the strength of the nonlinearity. The small magnitude asymmetry in offspring doubling times observed in unstressed \emph{Escherichia coli} generates a growth advantage that is only likely significant on evolutionary timescales ($>1000$ generations), but could be the maximum feasible based on diminishing returns for asymmetric partitioning.

Although protein aggregates have been identified to partition asymmetrically in \emph{Escherichia coli} and contribute to a fitness reduction under highly stressed conditions \cite{lindner2008asymmetric}, it is still unknown what subcellular compounds are being asymmetrically partitioned in unstressed conditions. As discussed by {\L}api\'nska et al. \cite{lapinska2019bacterial}, it is possible that asymmetry in offspring doubling times is a result of the cost of creating a symmetrical division in compounds such as ribosomes that have a positive fitness effect. While this has interesting implications on the potential evolution of asymmetric division, based on our results it is unlikely to explain the asymmetry observed in \emph{Escherichia coli}. Under this hypothesis, one would expect a high cost (in net doubling time) for a low amount of symmetry (compare with Figure~\ref{fig-Y}B).  Given that large asymmetry already corresponds with a large population growth advantage (Figure~\ref{fig-X}A), one would expect these two factors to work in concert to create equilibria with a much larger growth discrepancy.  Based on the natural assumption of diminishing returns with asymmetric division, the model presented here indicates that the partition of damaging compounds is a more likely source of the observed behavior.

\section*{Acknowledgments}
The authors are grateful to Roman Stocker and Martin Ackermann for helpful conversations and hospitality during the first author's sabbatical leave.

\bibliographystyle{alpha}
\bibliography{phenotypic_refs} 
\end{document}


\maketitle

\section{The combinatorial model}

\begin{figure}\centering\includegraphics[scale=1.4]{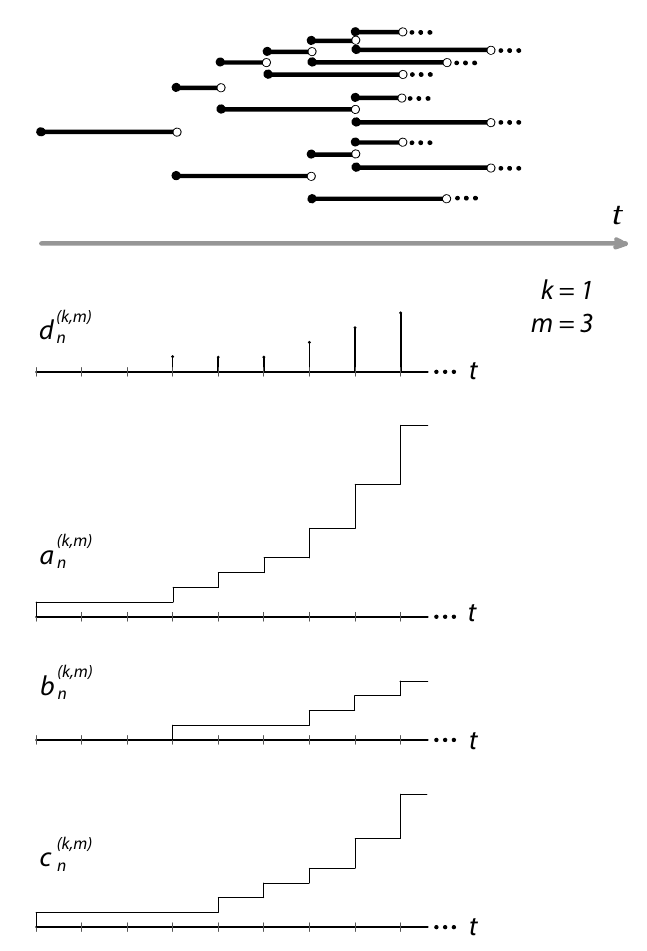}\caption{$(1,3)$-heterogeneous branching. Several corresponding variables from the main text are displayed below the branching diagram. Displayed is the number of divisions at time $t=n$ (denoted $d_n^{(k,m)}$), the  number of bacteria alive at time $t\in [n,n+1)$ (resp.\ $a_n^{(k,m)}$), the number of such bacteria with lifetime $k$ ($b_n^{(k,m)}$) and lifetime $m$ (resp. $c_n^{(k,m)}$).}\label{fig-tree}\end{figure}

Fix some time increment $\delta>0$ and let $k$ and $m$ be non-negative integers with $k\leq m$. Consider the scenario where a bacterium is born at time 0 and undergoes its first division at time $\delta m$. Then, starting with this division and for all divisions thereafter, each bacterium in the system splits into a child that will undergo division at time $\delta k$ and a child that will undergo division at time $\delta m$. We term this the \emph{$(k,m)$-heterogenous branching} and denote by $a_n^{(k,m)}$,  for $n=0,1,2,\ldots$, the \emph{number of bacteria} present at time $t\in [n\delta,(n+1)\delta)$. 
The first few divisions occurring for $k=1,m=3$ are illustrated in Figure~\ref{fig-tree}.

Consider also an auxiliary quantity to the number of bacteria alive at a given time. Namely, denote by $d_n^{(k,m)}$ ($n=0,1,2,\ldots$) the number of simultaneous \emph{divisions} occurring in the $(k,m)$-heterogenous branching at time $\delta n$. (See Figure~\ref{fig-tree}.)  While the number of bacteria alive at time $t\geq 0$ is constant on each interval $[n\delta,(n+1)\delta)$, the number of divisions is zero outside of those times at which divisions may happen, namely, the times $t=\delta n \gcd(k,m)$ for $n\in \mathbb N$. We henceforth  assume $k$ and $m$ to be relatively prime, i.e. $\gcd(k,m)=1$, as any common factors can be absorbed by the time increment $\delta$. Furthermore, all times will be given in units of $\delta$ (e.g. `time $m$' will refer to time $\delta m$). Our first observation is the following.

\begin{lemma}
For $n\geq m+1$,
\begin{equation}d^{(k,m)}_n=d^{(k,m)}_{n-k}+d^{(k,m)}_{n-m},\label{eq-recurrence-d}\end{equation}
with $d^{(k,m)}_0=\ldots=d^{(k,m)}_{m-1}=0$ and $d^{(k,m)}_{m}=1$.\label{lemma-d}
\end{lemma}
\begin{proof}
By construction, the first division occurs at time $m$. A bacterium dividing at time $n$ was born from a division either at time $n-m$ or at time $n-k$. Conversely, each bacterium born at time $n-m$ engenders a child that will incur a division at time $n$ (namely, a child of lifetime $m$) and each bacterium born at time $n-k$ engenders a child that will incur a division at time $n$ (namely, a child of lifetime $k$).
\end{proof}

The sequence $(a_n^{(k,m)})_n$ can be now be characterized as follows.

\begin{theorem} Let $a_n^{(k,m)}$ denote the number of bacteria in the $(k,m)$-heterogenous branching $(1\leq k\leq m)$ at time $t\in [n\delta,(n+1)\delta)$. Then, $a_0^{(k,m)}=\ldots=a_{m-1}^{(k,m)}=1$, and for all $n\in\mathbb N$ with $n\geq m$,
\begin{equation}a_n^{(k,m)}=a_{n-k}^{(k,m)}+a_{n-m}^{(k,m)}.\label{eq-recurrence}\end{equation}
Equivalently, the formal power series generating function of $(a_n)_n$ is
\begin{equation}T(x):=\sum_{n=1}^\infty a_n^{(k,m)} x^n=\frac{1-x^k}{(1-x^k-x^m)(1-x)}.\label{eq-gf}\end{equation}
In closed form, the sequence equals
\begin{equation}a_n=\sum_{r}\frac{(1-r^k)}{(1-r)(mr^{m}+kr^{k})}\,\frac{1}{r^{n}},\label{eq-explicit}\end{equation}
where the sum is taken over all (complex) roots $r$ of 
\begin{equation}1-x^k-x^m=0.\tag{CE}\label{eq-poly}\end{equation}
For all $1\leq k<m$ ($k,m\in\mathbb N$), \eqref{eq-poly} has a unique positive root $r_0$. The root $r_0$ is simple, takes value in the interval $(1/\sqrt[k]{2},1/\sqrt[m]{2})$, and is of least magnitude, in the sense that $|r|\geq r_0$ for all roots $r$ of \eqref{eq-poly}. If $k$ and $m$ are relatively prime, the inequality is strict, i.e. $|r|> r_0$ for all roots $r\neq r_0$ of \eqref{eq-poly}. When $k$ and $m$ are relatively prime, it follows that for large $n$, 
\begin{equation}a_n^{(k,m)}\sim \frac{(1-r_0^k)}{(1-r_0)(mr_0^{m}+kr_0^{k})}\,r_0^{-n}.\label{eq-asymptotic}\end{equation}
\label{thm-phenotypic-main}
\end{theorem}
\begin{proof}
To show that the number of bacteria satisfies \eqref{eq-recurrence}, observe that each division creates two bacteria: one which `takes the place' of the mother bacterium, and one additional one. Hence, each division adds exactly one new bacterium to the tally. So, for all $n\in\mathbb N$, \begin{equation}a_n^{(k,m)}=1+\sum_{i=1}^n d_i^{(k,m)},\label{eq-a-sum-d}\end{equation}
where the addition of the $1$ accounts for the fact that a single bacterium is already present at time $0$. Then, 
\begin{eqnarray*}a_{n-k}^{(k,m)}+a_{n-m}^{(k,m)}&=&2+\sum_{i=1}^{n-k} d_i^{(k,m)}+\sum_{i=1}^{n-m} d_i^{(k,m)}\\
&=&2+d_{1}^{(k,m)}+\ldots+d_{m-k}^{(k,m)}+\sum_{j=0}^{n-m-1}d_{n-k-j}^{(k,m)}+d_{n-m-j}^{(k,m)}\\
&=&2+d_{1}^{(k,m)}+\ldots+d_{m-k}^{(k,m)}+\sum_{j=0}^{n-m-1}d_{n-j}^{(k,m)}\\
&=&1+\sum_{j=0}^{n}d_{n-j}^{(k,m)}=a_{n-k}^{(k,m)},
\end{eqnarray*}
where we used the recurrence \eqref{eq-recurrence-d} and the fact that $d_{0}^{(k,m)}=\ldots=d_{m-1}^{(k,m)}=0$ while $d_{m}^{(k,m)}=1$ (Lemma~\ref{lemma-d}).
The remaining claims can now be shown using standard methods in linear difference equations (see e.g. \cite{Levesque1985}).
\end{proof}

\begin{remark}
In the asymptotic expression \eqref{eq-asymptotic}, since $r_0\in(1/2,1)$, the growth rate $-n\log(r_0)$ is strictly positive, as is to be expected. Furthermore, as can be inferred from \eqref{eq-asymptotic}, when $k$ and $m$ are relatively prime, the ratio of consecutive terms, namely ${a_n^{(k,m)}}/{a_{n+1}^{(k,m)}}$, converges as $n\to\infty$ to $r_0$.
\end{remark}

\begin{corollary}
Let $1\leq k< m$ and denote by $r_0$ the unique positive root of \eqref{eq-poly}. Then,
\begin{equation}
\frac{1}{m}<\frac{2}{k+m}<-\log_2(r_0)<\frac{1}{k}.
\end{equation}
Hence, for large $n$,
$a_{n}^{(k,m)}\gg a_{n}^{((k+m)/2,(k+m)/2)}$.\label{cor-homogeneity-growth}
\end{corollary}

\section{Composition of the population at time $n$}

We are next interested in the composition of the bacterial population at time $n$ as differentiated by the lifetime. Namely, let $b^{(k,m)}_n$ $(n=0,1,2,\ldots)$ denote the number of bacteria alive at time $t\in [n,(n+1))$ having \emph{lifetime $k$}. Similarly, let $c^{(k,m)}_n$ denote the number of bacteria alive at time $t\in [n,(n+1))$ having \emph{lifetime $m$}. (See Figure~\ref{fig-tree}.)
Both sequences satisfy the same recurrence as $(a^{(k,m)}_n)_n$ and $(d^{(k,m)}_n)_n$, starting from a different initial condition. In particular, we have the following.

\begin{theorem}
 Fix $k,m\in\mathbb N$ with $k<m$. Then, 
 $$b^{(k,m)}_0=\ldots=b^{(k,m)}_{m-1}=0\quad \text{and} \quad b^{(k,m)}_{m}=\ldots=b^{(k,m)}_{m+k-1}=1,$$
$$c^{(k,m)}_0=\ldots=c^{(k,m)}_{m+k-1}=1,$$
and, for $n\geq k+m$,
\begin{eqnarray}b^{(k,m)}_n&=&b^{(k,m)}_{n-k}+b^{(k,m)}_{n-m},\label{eq-recurrence-b}\\
c^{(k,m)}_n&=&c^{(k,m)}_{n-k}+c^{(k,m)}_{n-m}.\label{eq-recurrence-c}
\end{eqnarray}
\label{thm-bc}
\end{theorem}
\begin{proof}
By construction, the unique mother bacterium existing at time $0$ has lifetime $m$. Moreover, the first division, occurring at time $m$, produces a bacterium with lifetime $k$ and a bacterium with lifetime $m$.  Hence, $b^{(k,m)}_0=\ldots=b^{(k,m)}_{m-1}=0$, $b^{(k,m)}_{m}=\ldots=b^{(k,m)}_{m+k-1}=1,$ and $c^{(k,m)}_0=\ldots=c^{(k,m)}_{m+k-1}=1.$ Starting with the time increment $n=m+k$, each division produces exactly one bacterium with lifetime $k$ and one with lifetime $m$. Each time a bacterium of lifetime $k$ divides, it gets `replaced' by a bacterium of lifetime $k$ and, hence, does not increase the total number of bacteria with lifetime $k$. In contrast, each time a bacterium of lifetime $m$ divides, it creates a new bacterium of lifetime $k$, increasing the total of bacteria with lifetime $k$ by one for each such division. Let $e_n^{(k,m)}$ denote the number of bacteria with lifetime $m$ that are incurring division at time $n$. Observe that, by construction, $e_0^{(k,m)}=\ldots=e_{m-1}^{(k,m)}=0$ and $e_m^{(k,m)}=1$. Thereafter, $e_n^{(k,m)}=d_{n-m}^{(k,m)}$, as each bacterium with lifetime $m$ incurring division at time $n$ must have been born at time $n-m$ and, conversely, each division at time $n-m$ engendered exactly one child of lifetime $m$. By \eqref{eq-recurrence-d},
$$e_n^{(k,m)}=e_{n-k}^{(k,m)}+e_{n-m}^{(k,m)}.$$
Since $b_n^{(k,m)}=\sum_{i=1}^n e_i^{(k,m)}$, \eqref{eq-recurrence-b} holds by an analogous argument to that in the proof of Theorem~\ref{thm-phenotypic-main}. Since $c_n^{(k,m)}=a_n^{(k,m)}-b_n^{(k,m)},$ \eqref{eq-recurrence-c} follows by \eqref{eq-recurrence} and \eqref{eq-recurrence-b}.
\end{proof}

\begin{corollary}
For $n\geq m$, $b_n^{(k,m)}=a_{n-m}^{(k,m)}$. Furthermore, for $n\geq k$, $c_n^{(k,m)}=a_{n-k}^{(k,m)}$.
\label{cor-bc-eq-a}
\end{corollary}
\begin{proof}
Since  $b^{(k,m)}_0=\ldots=b^{(k,m)}_{m-1}=0$ and $b^{(k,m)}_{m}=\ldots=b^{(k,m)}_{m+k-1}=1$, by \eqref{eq-recurrence-b}, we have that, in fact, $b^{(k,m)}_{m}=\ldots=b^{(k,m)}_{2m-1}=1$. Comparison with the recurrence \eqref{eq-recurrence} and the initial conditions in Theorem~\ref{thm-phenotypic-main} yields that $b_n^{(k,m)}=a_{n-m}^{(k,m)}$. Similarly, since $c^{(k,m)}_k=\ldots=c^{(k,m)}_{m+k-1}=1$, by \eqref{eq-recurrence-c} we obtain that $c_n^{(k,m)}=a_{n-k}^{(k,m)}$.
\end{proof}

Combining the above corollary with Theorem~\ref{thm-phenotypic-main} allows us to recover the asymptotics of the two sequences.

\begin{corollary} 
For $n\geq k$, 
\begin{eqnarray}b^{(k,m)}_n&=&\sum_{r}\frac{r^m(1-r^k)}{(mr^{m}+kr^{k})(1-r)}\,\frac{1}{r^{n}}=\sum_{r}\frac{(1-r^k)^2}{(mr^{m}+kr^{k})(1-r)}\,\frac{1}{r^{n}},\label{eq-explicit-b}\\
c^{(k,m)}_n&=&\sum_{r}\frac{r^k(1-r^k)}{(mr^{m}+kr^{k})(1-r)}\,\frac{1}{r^{n}}=\sum_{r}\frac{r^{m+k}}{(mr^{m}+kr^{k})(1-r)}\,\frac{1}{r^{n}}.\label{eq-explicit-c}
\end{eqnarray}
where the sum is taken over all (complex) roots $r$ of $1-x^k-x^m=0$. When $k$ and $m$ are relatively prime, for large $n$, 
\begin{eqnarray}b_n^{(k,m)}&\sim& \frac{r_0^m(1-r_0^k)}{(mr_0^{m}+kr_0^{k})(1-r_0)}\,r_0^{-n}=\frac{(1-r_0^k)^2}{(mr_0^{m}+kr_0^{k})(1-r_0)}\,r_0^{-n},\label{eq-asymptotic-b}\\
 c_n^{(k,m)}&\sim& \frac{r_0^k(1-r_0^k)}{(mr_0^{m}+kr_0^{k})(1-r_0)}\,r_0^{-n}=\frac{r_0^{k+m}}{(mr_0^{m}+kr_0^{k})(1-r_0)}\,r_0^{-n},\label{eq-asymptotic-c}
\end{eqnarray}
where $r_0$ is the unique positive root of $1-x^k-x^m=0$.
\label{thm-phenotypic-BC}
\end{corollary}

A natural question from the applications perspective concerns the relative proportions of bacteria, among all those alive at some specified (large) time $t$, which have lifetime $k$ (resp.\ $m$). 
\begin{corollary}
 When $k$ and $m$ are relatively prime, the ratios of the number of bacteria with lifetime $k$ (resp. $m$) to the total number of bacteria converges, in the limit of large time, to
\begin{equation}\lim_{n\to\infty}\frac{b_n^{(k,m)}}{a_n^{(k,m)}}=1-r_0^k=r_0^m,\quad\quad \lim_{n\to\infty}\frac{c_n^{(k,m)}}{a_n^{(k,m)}}=r_0^k,\label{eq-proprtions}\end{equation}
where $r_0$ is the unique positive root of $1-x^k-x^m=0$.\label{cor-proportions}
\end{corollary}

Observe that by \ref{cor-homogeneity-growth}, as the ratio $m/k$ approaches $1$, the root $r_0$ approaches $1/2$, and so the large time proportion of bacteria with lifetime $k$ (resp. lifetime $m$) becomes close to $1/2$, as expected. In general, since $k<m$, the proportion of bacteria with lifetime $k$ is between $0$ and $1/2$, with the lower bound approached by letting $k$ be small and $m$ large. 

\section{Heterogeneous division as a stochastic process}

The $(k,m)$-heterogeneous branching is `noiseless', that is, exact. Each division produces one child of lifetime  $k$ and one child of lifetime  $m$ without fault. We are interested in characterizing the effects of small amounts of variability on the lifetimes on the behavior of the root $r_0$.

For this, consider the following stochastic process. At time $0$, a bacterium is `born' and has a random lifetime (time to division) drawn from some cumulative distribution function (cdf) $F$. Upon division, the bacterium splits into two `children' whose lifetimes are independent of one another and independent of that of the mother bacterium. Unlike in the Bellman-Harris process \cite{bellman1948theory}, however, the children's lifetimes are not identically distributed. Rather, the `weaker child' has lifetime  according to some cdf $F_1$ with mean $k$, i.e.
$$\int_0^\infty x\,dF_1(x)=k,$$
while the lifetime of the `stronger child' is distributed according to some cdf $F_2$ with mean $m$. Thereafter, each division follows the same probabilistic description, with children's lifetimes drawn at random, independently of one another and of the mother bacterium, but distributed according to $F_1$ and $F_2$ respectively.

This heterogeneous branching process is sufficiently closely related to the Bellman-Harris process so that, with some care, it can be analyzed by drawing on classical arguments. The arguments presented in this section will be sketched out, but can be made rigorous without much difficulty.

Specifically consider the following variations on the previously described process: one in which the original mother bacterium (born at time zero) has lifetime distributed as $F=F_1$  and the other in which it has lifetime distributed as $F=F_2$. Let $N_1(t)$ be the number of bacteria alive at time $t$ in the first process and $N_2(t)$ that in the second process. Furthermore, denote by $M_1$ and $M_2$ the corresponding ensemble averages, i.e. $M_1(t)=\mathbb E(N_1(t))$ and $M_2(t)=\mathbb E(N_2(t))$.
By conditioning on the lifetime of the mother bacterium (see e.g. \cite{Harris}), we can deduce:
\begin{eqnarray}
M_1(t)&=&\int_t^\infty \,dF_1(\tau)+\int_0^t M_1(t-\tau)\,dF_1(\tau)+\int_0^t M_2(t-\tau)\,dF_1(\tau)\label{eq-branching-1}\\
M_2(t)&=&\int_t^\infty \,dF_2(\tau)+\int_0^t M_1(t-\tau)\,dF_2(\tau)+\int_0^t M_2(t-\tau)\,dF_2(\tau)\label{eq-branching-2}.
\end{eqnarray}
For example, focusing on the process $N_1(t)$ where the mother bacterium has lifetime distributed as $F_1$, observe that if the mother's lifetime exceeds $t$, there is only one bacterium present at time $t$. This is accounted for by the term $\int_t^\infty \,dF_1(\tau)$ in \eqref{eq-branching-1}. In contrast, if the mother's lifetime is some quantity $\tau<t$, the number of offspring is the sum of those arising from two processes: the one in which the weaker child plays the role of the new mother bacterium, accounting for the term $\int_0^t M_1(t-\tau)\,dF_1(\tau)$ in \eqref{eq-branching-1},  and that in which the stronger child becomes the new mother, giving the term $\int_0^t M_2(t-\tau)\,dF_1(\tau)$ in \eqref{eq-branching-1}. An analogous reasoning for the case when the lifetime of the original mother bacterium is $F_2$ gives \eqref{eq-branching-2}.

At large time, the process exhibits exponential growth. Namely, $M_1(t)\sim \beta_1 e^{\alpha_1 t}$ and $M_2(t)\sim \beta_2 e^{\alpha_2 t}$ for some fixed scalars $\alpha_1,\beta_1,\alpha_2,\beta_2$ ($\beta_1,\beta_2>0$). From \eqref{eq-branching-1} and \eqref{eq-branching-2}, these must satisfy:

\begin{eqnarray}
\beta_1 e^{\alpha_1 t}&=&\beta_1 e^{\alpha_1 t} \int_0^\infty e^{-\alpha_1 \tau}\,dF_1(\tau)+\beta_2 e^{\alpha_2 t}\int_0^\infty e^{-\alpha_2 \tau}\,dF_1(\tau)\label{eq-branching-11}\\
\beta_2 e^{\alpha_2 t}&=&\beta_1 e^{\alpha_1 t} \int_0^\infty e^{-\alpha_1 \tau}\,dF_2(\tau)+\beta_2 e^{\alpha_2 t}\int_0^\infty e^{-\alpha_2 \tau}\,dF_2(\tau)\label{eq-branching-22}.
\end{eqnarray}

Since the only difference between the processes $N_1(t)$ and $N_2(t)$ is the lifetime of the original mother bacterium, with the subsequent offspring distributed in an identical manner across the two processes, it can be deduced that $\alpha_1=\alpha_2$. (In other words, the initial delay only affects the pre-factors $\beta_1$ and $\beta_2$.) Therefore:
\begin{eqnarray}
1&=&\int_0^\infty e^{-\alpha \tau}\,dF_1(\tau)+\frac{\beta_2}{\beta_1}\int_0^\infty e^{-\alpha \tau}\,dF_1(\tau)\label{eq-branching-111}\\
1&=&\frac{\beta_1}{\beta_2} \int_0^\infty e^{-\alpha \tau}\,dF_2(\tau)+\int_0^\infty e^{-\alpha \tau}\,dF_2(\tau)\label{eq-branching-222}.
\end{eqnarray}
That is, eliminating the term $\beta_1/\beta_2$ by substitution, the growth rate $\alpha$ becomes the solution to the fixed-point equation
\begin{equation}
1-\int_0^\infty e^{-\alpha \tau}\,dF_1(\tau)-\int_0^\infty e^{-\alpha \tau}\,dF_2(\tau)=0.\tag{FP}
\label{eq-fp}
\end{equation}
(Since \eqref{eq-fp} is monotonic in $\alpha$, any solution will be unique. Note, furthermore, that $\alpha\neq 0$.) So far, we have examined the asymptotic growth of the means $M_1(t)$ and $M_2(t)$. By  computing the analogous estimates for the asymptotic growth of the second moments $\mathbb E(N_1^2(t))$ and $\mathbb E(N_2^2(t))$ one can show, by adapting the reasoning of e.g. \cite{Harris}, that under relatively mild regularity hypotheses\footnote{E.g. it would suffice for $F_1$ and $F_2$ to have a density.}  on the distribution functions $F_1$ and $F_2$, the asymptotic growth rate $\alpha$ is observed with probability 1. That is, the population will grow at rate $\alpha$ in almost every instance.
  
In particular, when the lifetimes are constant, \eqref{eq-fp} reduces to \eqref{eq-poly} and, indeed, the process is simply the combinatorial process analyzed previously. When, in contrast, the lifetimes are uniformly distributed on $[k-\epsilon,k+\epsilon]$ and $[m-\epsilon, m+\epsilon]$ (for some fixed $\epsilon>0$), i.e. the lifetimes are `maximally noisy', in the sense that they are as likely to be close to $k$ and $m$ as to any other value in this interval, \eqref{eq-fp} becomes
\begin{equation}
1+\frac{e^{-\alpha(k+\epsilon)}-e^{-\alpha(k-\epsilon)}}{2\alpha\epsilon}+\frac{e^{-\alpha(m+\epsilon)}-e^{-\alpha(m-\epsilon)}}{2\alpha\epsilon}=0
\label{eq-uniform}
\end{equation}
The function $g:(0,\infty)\times(0,\infty)\to \R$ given by 
\begin{equation}g(x,\epsilon)=1+\frac{e^{-\alpha(k+\epsilon)}-e^{-\alpha(k-\epsilon)}}{2\alpha\epsilon}+\frac{e^{-\alpha(m+\epsilon)}-e^{-\alpha(m-\epsilon)}}{2\alpha\epsilon}\end{equation}
is continuous. Furthermore, for any $x>0$, $\lim_{\epsilon\to 0} g(x,\epsilon)=1-x^k-x^m$. (In fact, the convergence is monotonic, so that the root $f(x,\epsilon)=0$ decreases to $r_0$ as $\epsilon\searrow 0$.) The combinatorial model is therefore not rigid, that is, small perturbations to the lifetimes will not have an outsize effect on the predicted growth.

\begin{figure}\centering
\includegraphics[scale=.8]{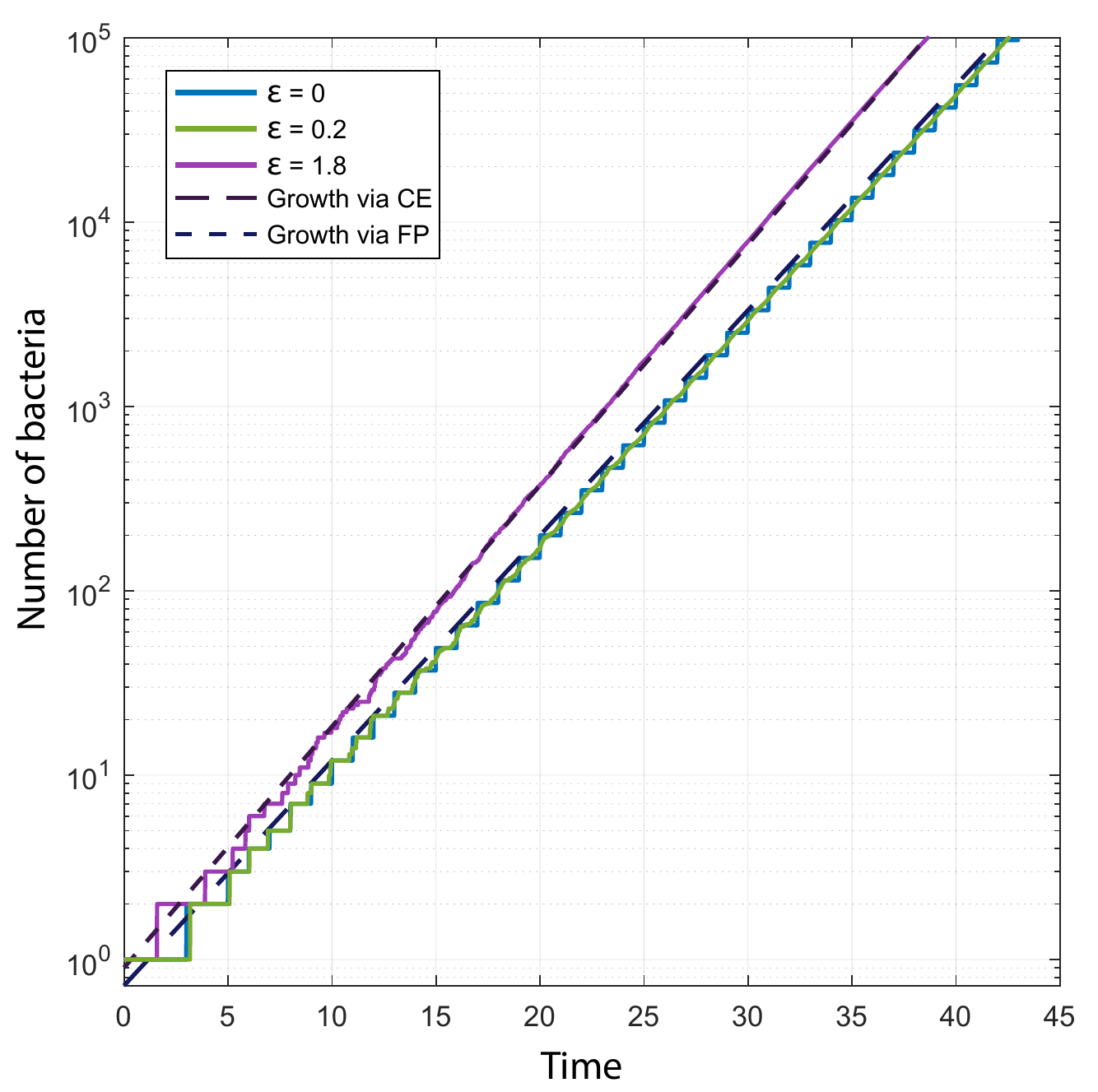}
\caption{$(2,3)$-heterogeneous branching with lifetimes uniformly distributed on $[2-\epsilon,2+\epsilon]$ and $[3-\epsilon,3+\epsilon]$. Plotted are two simulated sample paths of the process for $\epsilon=0.1$ (green) and $\epsilon = 1.8$ (purple), along with the corresponding asymptotics for the latter given by \eqref{eq-fp} (short dashed line). As $\epsilon$ gets smaller, the growth rate becomes accurately approximated by \eqref{eq-poly} (long dashed line). The deterministic version ($\epsilon = 0$) is plotted in blue.} 
\label{fig-example}
\end{figure}

We end this section with a worked example. Figure~\ref{fig-example} shows the number of bacteria alive at any given time in the combinatorial model for $(2,3)$-heterogeneous branching (blue), compared against the large-time asymptotic expression given in \eqref{eq-asymptotic} (dashed). The figure also shows a realization of the corresponding stochastic model with uniform noise of low variance ($\epsilon=0.2$, green) and with uniform noise of high variance ($\epsilon=1.8$, purple), along with their asymptotic growth rates as given in \eqref{eq-uniform} (dashed). For the non-arithmetic stochastic models, such as the uniform model with any nonzero $\epsilon$, the divisions will eventually desynchronize leading to a 'smoother' growth. We take this opportunity to emphasize that the combinatorial model is deterministic, while the stochastic model in the limit of large time grows exponentially with rate $\alpha$ given by \eqref{eq-fp} (specializing to \eqref{eq-uniform} in the uniform case) for almost every realization.

As predicted, the asymptotic population growth rate in the stochastic model with low variability is close to that of the combinatorial model. This is particularly useful from the computational perspective, as the positive root of \eqref{eq-poly} is generally considerably easier to numerically approximate than that of \eqref{eq-fp}.

\section{Analytic properties of $r_0$}
We are interested in behavior of $r_0$,  the unique positive root of the characteristic equation $1-x^k-x^m=0$, in the presence of `increasing heterogeneity'. Concretely, fix $\ell_0>0$ and consider the function $f:\left(-\ell_0,\ell_0\right) \to\R,$
given by 
\begin{equation}f(x)=\alpha, \text{ where }\alpha\text{ satisfies }1-2^{-\alpha(\ell_0-x)}-2^{-\alpha(\ell_0+x)}=0.\label{eq-def-f}\end{equation}
Note that $f$ is well defined. Indeed, fixing $p,q>0$ and considering the function $g:\R\to\R$, $g(x)=1-x^p-x^q$, we have that $g(0)=0$, $\lim_{x\to\infty} g(x)=-\infty$, and $g$ is strictly decreasing on $(0,\infty)$. Hence, $g$ has a unique root in $(0,\infty)$. 

Furthermore, $f$ is real analytic on its domain and therefore has the Taylor expansion about $0$ given by
\begin{equation}
 f(x)=f(0)+x f'(0)+\frac{x^2}{2}f''(0)+\frac{x^3}{6}f'''(0)+O(x^4).
\end{equation}
Computing the derivatives (by implicit differentiation of \eqref{eq-def-f}) yields
\begin{eqnarray}
 f(0)&=&\frac{1}{\ell_0}\\
 f'(0)&=&0\\
  f''(0)&=&\frac{\log(2)}{\ell_0^3}\label{eq-second-derivative}\\
  f'''(0)&=&0
\end{eqnarray}
Hence,
\begin{equation}
 f(x)=\frac{1}{\ell_0}+\frac{\log(2)}{2\ell_0^3}\,x^2+O(x^4).\label{eq-quadratic}
\end{equation}
Recall that when $k=m=(k+m)/2=\ell_0$, $2^{-1/\ell_0}$ is the unique positive root of $1-x^k-x^m=0$. When the lifetimes of the offspring are unequal, and are instead given by $k=\ell_0-x$ and $m=\ell_0+x$, the growth rate $f$ (equaling $-\log_2(r_0)$ for $r_0$ which appears in \eqref{eq-asymptotic}, \eqref{eq-asymptotic-b}, \eqref{eq-asymptotic-c}, and \eqref{eq-proprtions})  increases quadratically as a function of the `heterogeneity' $x$.

\bibliographystyle{alpha}
\bibliography{phenotypic_refs}